# Spin-orbit interaction in Au structures of various dimensionalities

Xiaoping Yang,[1, *] Jian Zhou,[1] Hongming Weng,[1] and Jinming Dong[1, †]

[1]*National Laboratory of Solid State Microstructures and Department of Physics, Nanjing University, Nanjing 210093, P. R. China*

## Abstract

Variation of the geometrical and electronic properties of the gold materials in different dimensions has been investigated by *ab initio* method, taking into account the spin-orbit interaction. It is found that spin-orbit effects in different dimensional Au materials depend greatly on fundamental symmetry and dimensionality. For single walled gold nanotubes, spin-orbit interaction decreases significantly the conducting channel number of achiral tube (4, 0), and leads to spin splitting at Fermi level of chiral tube, indicating that quasi-1D gold tube can be a good candidate for the spin-electron devices. Furthermore, our results suggest that cage cluster might be synthesizable experimentally by taking gold tube structure as parent material.

---

[*]Present address: Max-planck Institute for Solid Research, Heisenbergstr. 1, D-70569 Stuttgart, Germany;

Electronic address: `bunnyxp@hotmail.com`

[†]Corresponding author. Email address: `jdong@nju.edu.cn`

Theoretical and experimental researches on the nanomaterials and nanowires have been widely done in the world due to their fundamental importance and also tremendous potential applications in the nanotechnology. Among the coinage metals, gold occupies a unique position. For example, the Au nanowires have been extensively studied theoretically [1–7] since they were observed experimentally [8–11], among which the small diameter gold nanowire could keep only a single shell [8], called as the single-walled gold nanotube (SWGNT). Recently, they have been synthesized experimentally at 150 K in an UHV—TEM [10]. The thinnest SWGNT, 4 Å in diameter, was found to be the (5,3), which is composed of five atomic rows, coiled round the tube axis. In addition, the gold cluster $Au_{20}$ in a tetrahedral structure was also found recently by the photoelectron spectrum combined with the density functional calculation [12]. The most recent surprise in Au cluster chemistry is the prediction of highly stable (Magic number) $Au_{32}$ [13] and $Au_{50}$ [14] cage clusters, especially $Au_{32}$ has the same icosahedral ($I_h$) symmetry as $C_{60}$. It is well known that the dramatic changes of the chemical and physical properties of a material can happen when its size becomes smaller and smaller, transforming from the bulk into the nano-scale levels, which makes the material exhibit a lot of fascinating and unconventional properties. This greatly motivated us to explore the behaviors of the gold materials in different dimensions, e.g., from three-dimensional bulk gold down to low dimensional SWGNT and cluster, with the emphasis on the variation of their geometrical and electronic properties over their dimensionality.

It is well known that the relativistic effect is important in the heavy elements, like the gold, which contains the scalar-relativistic (SR) one and the spin-orbit (SO) interaction. Usually, the former is much larger than the latter and plays a decisive role in the correct geometrical structures of the ground state of gold materials [15]. For example, the lattice constant of bulk-Au can be corrected from 4.286 Å to 4.094 Å (experimental data, 4.078 Å) [16] by the SR effect. And for SWGNT, we find the bond length of (4,0) gold tube contracts by about 0.2 Å due to the SR effect, and the SO interaction only induce a contract of about 0.01 Å. Even thus, the SO interaction has still been demonstrated to have a significant effect on the electronic properties in the different dimensional Au materials, which is just objective of this paper.

We carried out our numerical calculations by using the Vienna *ab initio* Simulation Package (VASP) within the framework of local density approximation (LDA) [17], which is a plane wave method with the SR correction. The ion-electron interaction was modeled by the projector augmented wave (PAW) method [18] with a uniform energy cutoff of 500 eV. The smearing width was taken to be 0.04 eV in the ground state. The supercell geometry for the SWGNT and cluster has been used, in which the tubes and cluster are aligned in a hexagonal array with the closest distance between the adjacent tubes or clusters being 24 Å, larger enough to prevent the interactions between neighboring tubes or clusters. Using the same computational method, we have calculated the band structure of the SWGNT (8, 4) plus one atomic row at the center of the tube, and the obtained result is well consistent with that in Ref. [1].

Here, we choose chiral SWGNT (5, 3) and achrial SWGNTs (4,0), (6,3) as examples of the gold tube structure; choose $Au_{20}$ and $Au_{32}$ as examples of the gold cluster structure. In Fig. 1 shown are their structures. The changes of the bond length, the binding energy $E_b$, $E_b^{so}$, and the binding energy difference $\triangle E_b$ in the different dimensional gold materials are calculated by including or excluding the SO coupling, in order to study the role of SO interaction. The results are given in Table I.

For bulk structure, our Au-Au bond length of 2.884 Å, which is well consistent with the experimental value of 2.88 Å [19]. The SO interaction does not induce bond length change, which is also in agreement with a previous FPLAPW calculation [20], but increases the binding energy by 0.141 eV/atom. On the other hand, spin-orbit coupling energy is found to be 0.202 eV from our atomic calculations with and without SO coupling. In Table I, it can be found that SO interaction induces the largest bond length changes of -0.01 (-0.013 Å) for SWGNT (cluster), indicating that the bond length shrinking is enhanced as the dimensionality decreases [∼ 0.36% for SWGNT and ∼ 0.46% for cluster, respectively]. The binding energy decreases with the decreasing dimensionality. SO coupling increases the binding energy in the these material. However, the binding energy difference $\triangle E_b$ decreases as the dimensionality decreases, because the ratio of the bond energy to the total energy decreases with the dimensionality decrease. For all SWGNTs, we find that the binding energy differences $\triangle E_b$ are always close to that of the Au (111) plane (an unsupported monolayer), being 0.127 eV/atom, within a variation scope of about ±0.002 eV/atom. Finally, for cluster we find $E_b$, $E_b^{so}$ and $\triangle E_b$ of cage $Au_{32}$ is obviously larger than those of $Au_{20}$, and much close to those of SWGNT [especially (5,3) with 38 atoms]. It may due to the hollow structure and the quasi-2D surface of $Au_{32}$, and indicates the possible synthesis experimentally by taking gold tube structure as parent material.

The density of states (DOSs) and the energy band structures of bulk, SWGNT (5,3), and cluster $Au_{32}$ are shown in Fig. 2 and 3, respectively. The outermost orbit of Au is $5d^{10}6s^1$, so in each case the central main structure of the valence band is due to the $d$ states, whereas the charges outside its shoulders have significant $s$ character due to $s$-$d$ hybridization. For example, the shaded area in Fig. 2 is the projected $s$ component from calculations without SO interaction. With the dimension lowering, it is found that the band structures become less dispersive enhancing the related DOS, and $d$ band edge rises progressively, no matter the SO effect is included or not. It suggests that interatomic interactions become much stronger in low dimensional Au materials. After including SO coupling, the $d$ bandwidths is increased in the same dimensional system (see Fig. 2), indicating the further enhancement of $s$-$d$ hybridization.

In the band structure of 3D bulk structure, there exist some degenerate energy bands due to the system's symmetry (including the inversion symmetry). The SO interaction partially removes the $d$ bands' degeneracy. However, the states near the Fermi Level are not affected, since they are mainly of s character. Meantime, in Figs. 3a and 3d the spin degeneracy is not lifted. On the other hand, the chiral SWGNT (5,3) has no the inversion symmetry, but possesses a spiral symmetry. In contrast to the 3D bulk structure, its energy bands near the Fermi level are only doubly degenerate in spin, and SO interaction lifts spin degeneracies of all energy bands near Fermi level, inducing the spin splitting, which leads to the complete separation of spin conduction channels (see Figs. 3b and 3e). Thus, the SO interaction produces an energy splitting between those states with different spin orientations in chiral SWGNT. In the above two kinds of gold structures, the bands crossing Fermi level come from different symmetries, and so they would respond differently to the SO interaction. This is appreciable in Figs. 2 and 3 since the SO splitting is band-dependent. All those can be understood by Kramers' theorem on time reversal symmetry, which states that $E_{\vec{k},\sigma} = E_{-\vec{k},-\sigma}$, where $E_{\vec{k},\sigma}$ is the energy of the eigenstate with wave vector $\vec{k}$ and spin $\sigma$. But if a system has also a space inversion symmetry, making $E_{\vec{k},\sigma} = E_{-\vec{k},\sigma}$, then $E_{\vec{k},\sigma} = E_{\vec{k},-\sigma}$, i.e., spin degeneracy cannot be removed. For the chiral SWGNTs, due to lack of space inversion symmetry, only the first condition ($E_{\vec{k},\sigma} = E_{-\vec{k},-\sigma}$) holds, therefore their spin degeneracy can be lifted. This causes an asymmetry in momentum space for the spin up and down energy in the chiral tubes. And the motions of electrons with spin up and spin down are completely decoupled, indicating that quasi-1D SWGNT can be good candidates for spin-electron devices.

Finally, for the energy levels of cluster $Au_{32}$, our analysis shows that SO interaction makes HOMO-LUMO gap reduced by about 0.59 eV, and the spin degeneracy still holds although the inversion center is absent in cluster. Since removal of Kramers' degeneracy results from the movement of electrons in an electric field, it can not be expected to occur in a finite cluster. HOMO-LUMO gap of $Au_{20}$ is also decreased by 0.47 eV. Our analysis for binding energy and electronic structure of cluster indicates that the quasi 2D surface seems to strengthen the SO effect on hollow cage cluster, which reveals the complexity and importance of SO coupling in cluster.

In above paragraph, we have discussed the change of band structures of chiral SWGNT (5, 3). Now in Fig. 4, we also show the energy band structures of achiral SWGNTs (4,0) and (6,3). The two tubes have the space inversion symmetry, and so their spin degeneracy always remains no matter the SO effect is included or not. It is different from the chrial SWGNTs, as illustrated in Fig. 4, Figs. 3b and 3e. Since achiral SWGNTs have the higher spiral symmetry, the degenerate energy bands exist in their band structures. In Fig. 4, we can find the obvious splitting of degenerate $s$-bands near Fermi Level due to the SO interaction and enhanced $s$-$d$ hybridization. The most striking change is the band splitting of 0.22 eV at $\Gamma$ point near the Fermi level of (4,0) tube, leading to a decrease of the conduction channel number from original five to four.

In summary, we study the structural and electronic properties of the different dimensional Au materials: from bulk structure to SWGNTs, then to cluster. Our researches indicate that SO-related effects in different dimensional Au materials depend greatly on their symmetry and dimensionality. SO coupling plays an important role in electronic property of Au materials. Specially, the effect of SO on the electronic property is stronger in low dimensional structure than that in 3D bulk structure. Reason is the $d$ band is much closer to the Fermi Level in low dimensional system, and $s$-$d$ hybridization can be further strengthened by including the SO coupling. The SO interaction removes spin degeneracy in chiral SWGNTs, and causes a decrease of the conduction channel number around the Fermi Level for (4,0) SWGNT. Finally, HOMO-LUMO gap of cluster is considerably reduced by SO coupling. The binding energy of cage structure is found to be much close to that of tube structure. Cage cluster could be expected to be candidate material experimentally based on parent material SWGNTs.

Acknowledgments: This work was supported by the Natural Science Foundation of China under Grant No. 10474035 and 90503012, and also by the State Key program of China through Grant No. 2004CB619004.

# TABLE

TABLE I: For different dimensional gold structures, the bond length $a$, $b$, $c$ and $d$, defined as in Fig. 1, are calculated without SO interaction. The numbers in parentheses are the corresponding change after including the SO interaction. Binding energy $E_b$ (without SO coupling), $E_b^{so}$ (with SO coupling), and their difference $\triangle E_b$ are also shown in Table in the unit of eV/atom. In addition, spin-orbit coupling energy is found to be 0.202 eV from our atomic calculations with and without SO coupling.

| Au-material | $a$(Å) | $b$(Å) | $c$(Å) | $d$(Å) | $E_b$ | $E_b^{so}$ | $\triangle E_b$ |
|---|---|---|---|---|---|---|---|
| bulk | 2.884(0.0) | | | | 4.2834 | 4.4239 | 0.141 |
| SWGT(6,3) | 2.724(-0.010) | 2.724(-0.010) | 2.665(-0.004) | | 3.4759 | 3.6051 | 0.129 |
| SWGT(5,3) | 2.744(-0.010) | 2.754(-0.010) | 2.65(-0.005) | | 3.3654 | 3.4948 | 0.129 |
| SWGT(4,0) | 2.705(-0.010) | 2.705(-0.010) | 2.777(-0.002) | | 3.2963 | 3.4235 | 0.127 |
| cluster $Au_{32}$ | 2.675(-0.009) | 2.754(-0.006) | | | 3.3505 | 3.4696 | 0.119 |
| cluster $Au_{20}$ | 2.624(-0.003) | 2.640(-0.007) | 2.744(-0.008) | 2.853(-0.013) | 3.1826 | 3.2906 | 0.108 |

**Figure Captions**

Fig. 1 (Color online) Geometrical structure of (a) SWGNT (5,3), (b) SWGNT (4,0), (c) SWGNT (6,3), (d) cluster $Au_{20}$, and (e) cluster $Au_{32}$.

Fig. 2 (Color online) calculated density of States around Fermi Level of the different Au materials: (a) bulk structure, (b) SWGNT (5,3), (c) cluster $Au_{32}$. Black dotted (red solid) curves represent the calculations without (with) SO interaction, and the shaded area is the projected $s$ component from calculations without SO interaction. The Fermi level is set at zero.

Fig. 3 (Color online) Calculated band structures around Fermi Level of the different Au materials: bulk structure (a) without and (d) with SO interaction; SWGNT (5,3) (b) without and (e) with SO interaction; cluster $Au_{32}$ (c) without and (f) with SO interaction. Red numbers label the degenerate states of different energies without SO interaction. The Fermi level is set at zero.

Fig. 4 (Color online) Band structures along the tube symmetry axis $\Gamma X$ of SWGNT (4,0): (a) without and (c) with SO interaction, and SWGNT (6,3) (b) without and (d) with SO interaction. Red numbers label the degenerate states of different energies without SO interaction. The Fermi level is set at zero.

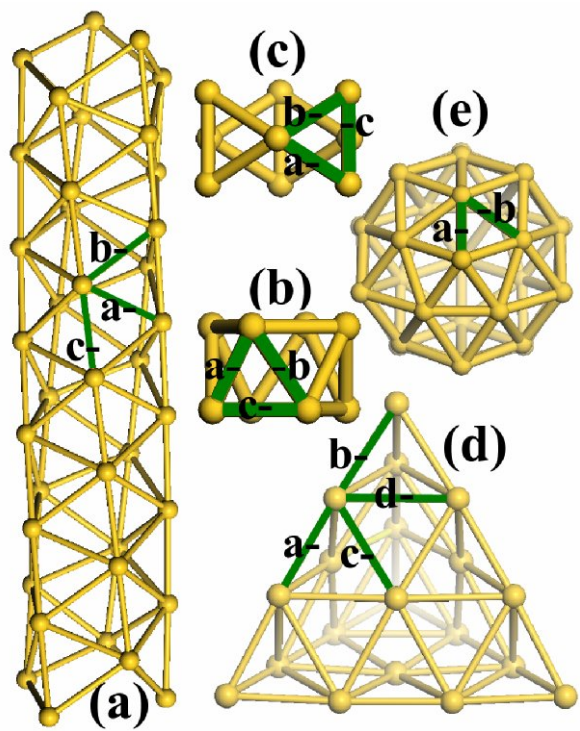

**Figure 1**

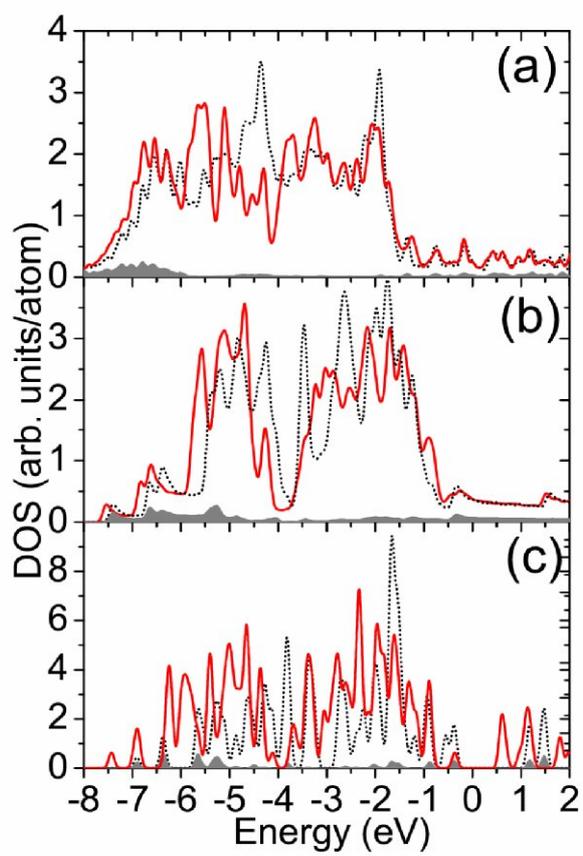

**Figure 2**

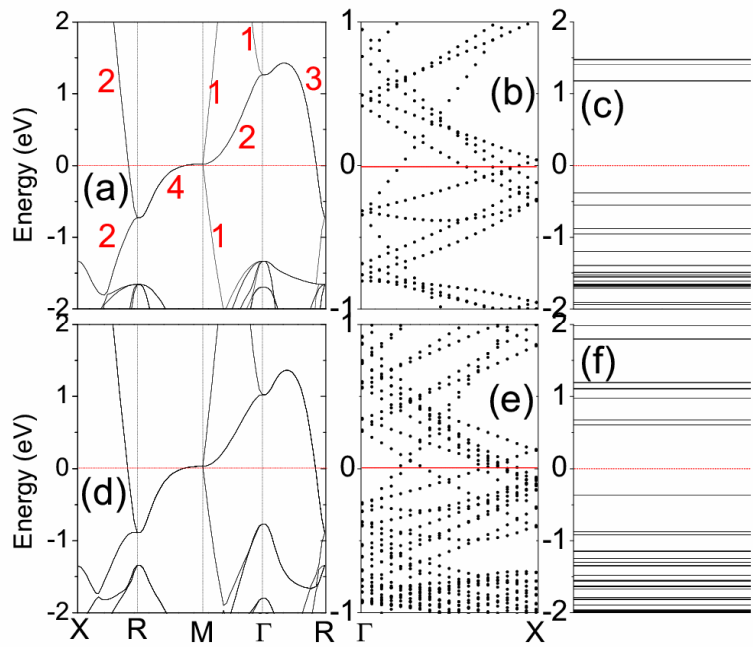

**Figure 3**

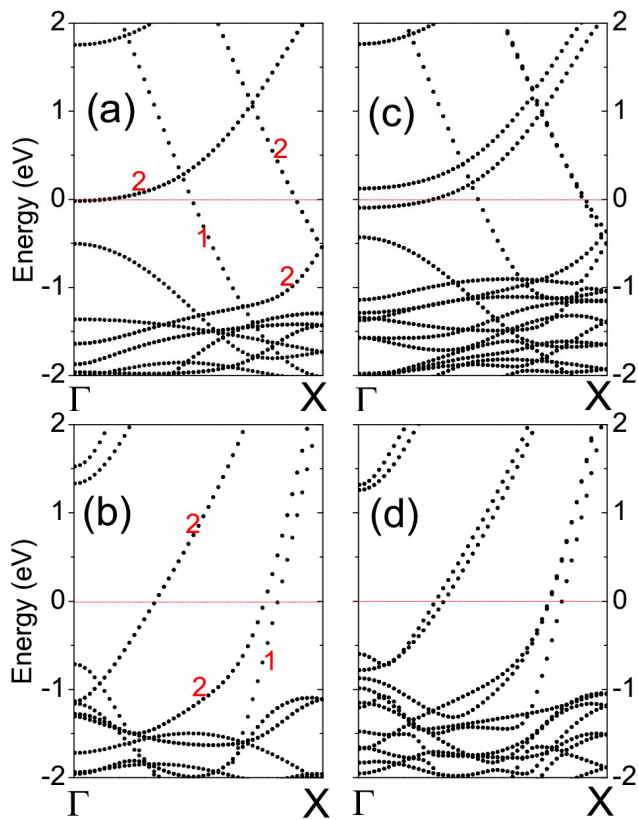

**Figure 4**